\documentstyle[12pt]{article}
\def\ie{{\em i.e.}}
\def\eg{{\em e.g.}}

\def\beq{\begin{equation}}
\def\eeq{\end{equation}}

\catcode`\@=11 
\def\coeff#1#2{{\textstyle{#1\over #2}}}

\def\vev#1{\langle #1\rangle}
\def\lsim{\mathrel{\mathpalette\@versim<}}
\def\gsim{\mathrel{\mathpalette\@versim>}}
\def\@versim#1#2{\vcenter{\offinterlineskip
    \ialign{$\m@th#1\hfil##\hfil$\crcr#2\crcr\sim\crcr } }}
\def\etal{{\em et. al.}}
\def\JL{J. L. Lopez}
\def\DVN{D. V. Nanopoulos}

\def\r#1{$\bf#1$}
\def\rb#1{$\bf\overline{#1}$}

\def\t1{{\tilde 1}}

\def\GeV{\,{\rm GeV}}
\def\TeV{\,{\rm TeV}}

\def\y{\,{\rm y}}

\def\to{\rightarrow}

\def\NPB#1#2#3{Nucl. Phys. B {\bf#1} (19#2) #3}
\def\PLB#1#2#3{Phys. Lett. B {\bf#1} (19#2) #3}

\def\PRD#1#2#3{Phys. Rev. D {\bf#1} (19#2) #3}
\def\PRL#1#2#3{Phys. Rev. Lett. {\bf#1} (19#2) #3}

\def\MODA#1#2#3{Mod. Phys. Lett. A {\bf#1} (19#2) #3}
\def\IJMP#1#2#3{Int. J. Mod. Phys. A {\bf#1} (19#2) #3}
\def\TAMU#1{CTP-TAMU-#1}

\def\hepph#1{({\tt hep-ph/#1})}

\begin{document}
\begin{flushright}
\baselineskip=12pt
CERN-TH-95/260\\
DOE/ER/40717--17\\
CTP-TAMU-39/95\\
ACT-14/95\\
\tt hep-ph/9510246
\end{flushright}

\begin{center}
\vglue 1cm
{\Large\bf Lowering $\alpha_s$ by flipping SU(5)\\}
\vglue 0.75cm
{\Large John Ellis,$^1$ Jorge L. Lopez,$^2$ and D.V. Nanopoulos$^{3,4}$\\}
\vglue 0.5cm
\begin{flushleft}
$^1$CERN Theory Division, 1211 Geneva 23, Switzerland\\
$^2$Department of Physics, Bonner Nuclear Lab, Rice University\\ 6100 Main
Street, Houston, TX 77005, USA\\
$^3$Center for Theoretical Physics, Department of Physics, Texas A\&M
University\\ College Station, TX 77843--4242, USA\\
$^4$Astroparticle Physics Group, Houston Advanced Research Center (HARC)\\
The Mitchell Campus, The Woodlands, TX 77381, USA\\
\end{flushleft}
\end{center}

\vglue 0.5cm
\begin{abstract}
We show that the prediction for $\alpha_s(M_Z)$ in flipped SU(5) is naturally
lower than in minimal SU(5), and that the former can accommodate
the full range of $\alpha_s(M_Z)$ presently allowed by experiment. Our
computations include two-loop ($\delta_{\rm 2loop}$) and light ($\delta_{\rm
light}$) and heavy ($\delta_{\rm heavy}$) threshold effects. Unlike minimal
SU(5), in flipped SU(5) the heavy threshold effects can naturally decrease
the predicted value of $\alpha_s(M_Z)$. We also show that the value of the
proton lifetime into the dominant channel $p\to e^+\pi^0$ is within the
observable range at SuperKamiokande, and should discriminate
against minimal supersymmetric
SU(5), where the dominant mode is $p\to\bar\nu K^+$.
\end{abstract}
\vspace{1cm}
\begin{flushleft}
\baselineskip=12pt
CERN-TH-95/260\\
DOE/ER/40717--17\\
CTP-TAMU-39/95\\
ACT-14/95\\
October 1995
\end{flushleft}

\vfill\eject
\setcounter{page}{1}
\pagestyle{plain}
\baselineskip=14pt

One of the most impressive pieces of circumstantial evidence for
Grand Unified Theories is the successful correlation they yield
between $\alpha_s(M_Z)$ and $\sin^2\theta_W$ \cite{GUTs}. Historically, this
was first phrased as a prediction of $\sin^2\theta_W$ based on
the measured value of $\alpha_s$, whose qualitative agreement
with early electroweak data was impressive, particularly for
supersymmetric GUTs \cite{amaldi-costa}. The advent of precision electroweak
data from LEP and elsewhere \cite{lepewwg}, and the measurement of the
top-quark mass $m_t$ \cite{cdfd0} has enabled $\sin^2\theta_W$ to be determined
with such accuracy that it is natural to turn the GUT correlation round the
other way, and use it to predict $\alpha_s(M_Z)$, which is still
not known with satisfactory precision \cite{bethke}. The prediction of a
minimal non-supersymmetric SU(5) GUT is disastrously low, and
debate centres on the the possibility of discriminating between
different supersymmetric GUTs and/or obtaining indirect
constraints on the scale of supersymmetry breaking.\\

Progress on these questions is hampered by the persistent
imprecision in the experimental value of
$\alpha_s(M_Z)$, which is usually quoted as $0.118 \pm 0.006$.
Most determinations lie within one standard deviation of this
mean, but this consistency hides two schools of thought: one
supported more strongly by high-energy determinations of $\alpha_s$
which favour $\alpha_s(M_Z) \ge 0.120$ \cite{lepewwg},
and one favoured more strongly by low-energy determinations of $\alpha_s$ which
favour $\alpha_s(M_Z) \le 0.115$ when evolved up to a higher scale
\cite{lowguys}. It is not clear whether this apparent discrepancy is real, but
it has led to various supersymmetric speculations, including the possibility
that gluinos are sufficiently light to alter the evolution of $\alpha_s$
between low energies and $M_Z$ \cite{LightGluino}, or that radiative
corrections due to other light sparticles such as the chargino and top-squark
reduce the value of $\alpha_s(M_Z)$ inferred from LEP data \cite{KSW}.\\

The minimal supersymmetric SU(5) prediction of $\alpha_s(M_Z)$
definitely lies on the upper side of the experimental range,
and even above it, with values around $0.130$ or higher being
preferred \cite{smooth,baggeretal}. This preference should
however be treated with caution, as the predicted value may be reduced by GUT
threshold effects \cite{EKNIII,heavy} or by `slop' induced by Planck-scale
interactions \cite{planck}. One example of a supersymmetric GUT which makes a
lower prediction for $\alpha_s(M_Z)$ is the Missing-Doublet Model (MDM)
\cite{MPM}, in which the GUT threshold corrections are generally negative
\cite{MDM}. However, the MDM is quite cumbersome, containing several large
Higgs representations that cannot easily be accommodated in a string
framework \cite{elnhigherlevel}.\\

By far the most economical realization of the missing-partner mechanism is that
in flipped SU(5) \cite{oldflip}, which needs only \r{10} and \rb{10}
representations of GUT Higgses and has been derived from string
\cite{stringflip}. This model also provides another natural
mechanism for reducing the prediction for $\alpha_s(M_Z)$,
namely if the scale at which $\alpha_s$ and the SU(2) electroweak
coupling become equal is lower than the scale at which the SU(5)
and U(1) couplings become equal \cite{EKNII}.\\

In this paper we make a quantitative exploration of this possibility in the
supersymmetric minimal flipped SU(5) GUT \cite{oldflip}. In addition to
exploring the above-mentioned possible difference in unification scales, we
also calculate for the first time the GUT threshold effects in flipped SU(5),
finding that they may be naturally negative. We also demonstrate that the
flipped SU(5) mechanism for reducing $\alpha_s(M_Z)$ has an observable
signature if it lies in the lower part of the range currently allowed by
experiment \cite{bethke}, namely that proton decay into $e^+\pi^0$
should be observable in the SuperKamiokande detector.\\

The starting point for our discussion is
the lowest-order prediction for $\alpha_s(M_Z)$ in SU(5) GUTs, namely
\begin{equation}
\alpha_s(M_Z)={\coeff{7}{3}\,\alpha\over 5\sin^2\theta_W-1}\ ,
\label{eq:LO}
\end{equation}
which
yields the present central experimental value of
$\alpha_s(M_Z)\approx0.118$ for $\sin^2\theta_W=0.231$ and
$\alpha^{-1}=128$. It is possible to include
two-loop corrections ($\delta_{\rm 2loop}$), and light
($\delta_{\rm light}$) and heavy ($\delta_{\rm heavy}$)
threshold corrections by making the following
substitution in Eq.~(\ref{eq:LO}) \cite{EKNII}:
\begin{equation}
\sin^2\theta_W\to \sin^2\theta_W-\delta_{\rm 2loop}
-\delta_{\rm light}-\delta_{\rm heavy}\ .
\label{eq:NLO}
\end{equation}
Here $\delta_{\rm 2loop}\approx0.0030$
whereas $\delta_{\rm light}$ and
$\delta_{\rm heavy}$ can have either sign.
Neglecting $\delta_{\rm light}$ and
$\delta_{\rm heavy}$, the SU(5) prediction increases to
$\alpha_s(M_Z)\approx0.130$. Therefore, if one wishes to
obtain a value of $\alpha_s(M_Z)$ within one standard deviation
of the present central experimental value,
one requires a non-negligible
contribution from $\delta_{\rm light}$ and/or $\delta_{\rm heavy}$,
so that the combined correction ($\delta_{\rm 2loop}
+\delta_{\rm light}+\delta_{\rm heavy}$) is suppressed. In large
regions of parameter space, $\delta_{\rm light}>0$ and does not help. As we
discuss in more detail later, exploiting $\delta_{\rm heavy}$ is difficult in
minimal SU(5) because of proton decay constraints \cite{EKNIII,baggeretal},
unless one goes to the limit
$m_0\gg m_{1/2}$ and $m_0\sim1\TeV$ which suppresses proton decay. However,
even in this case the decrease in $\alpha_s(M_Z)$ is not large:
$\alpha_s(M_Z)>0.123$ \cite{baggeretal}.
One way to circumvent these problems
is the cumbersome missing doublet model (MDM) \cite{baggeretal,MDM}:
here we take the more elegant route of flipped SU(5).\\

In this model, there is a first unification scale $M_{32}$ at
which the SU(3) and SU(2) gauge couplings become equal,
given to lowest order by \cite{faspects}
\begin{eqnarray}
{1\over\alpha_2}-{1\over\alpha_5}&=&{b_2\over2\pi}\,
\ln{M_{32}\over M_Z}\ ,
\label{eq:RGE2}\\
{1\over\alpha_3}-{1\over\alpha_5}&=&{b_3\over2\pi}\,
\ln{M_{32}\over M_Z}\ ,
\label{eq:RGE3}
\end{eqnarray}
where $\alpha_2=\alpha/\sin^2\theta_W$, $\alpha_3=\alpha_s(M_Z)$, and the
one-loop beta functions are $b_2=+1$, $b_3=-3$. On the other hand, the
hypercharge gauge coupling $\alpha_Y={5\over3}
(\alpha/\cos^2\theta_W)$ evolves in general to a different
value $\alpha_1'$ at the scale $M_{32}$:
\begin{equation}
{1\over\alpha_Y}-{1\over\alpha_1'}={b_Y\over2\pi}\,
\ln{M_{32}\over M_Z}\ ,
\label{eq:RGEY}
\end{equation}
with $b_Y={33\over5}$. Above the scale $M_{32}$ the gauge group is
SU(5)$\times$U(1), with the U(1) gauge coupling $\alpha_1$ related to
$\alpha_1'$ and
the SU(5) gauge coupling ($\alpha_5$) by
\begin{equation}
{25\over\alpha_1'}={1\over\alpha_5}+{24\over\alpha_1}\ .
\label{eq:U(1)}
\end{equation}
The SU(5) and U(1) gauge couplings continue to
evolve above the scale $M_{32}$, eventually becoming
equal at a higher scale $M_{51}$. The consistency condition that
$M_{51}\ge M_{32}$ requires $\alpha_1(M_{32})\le\alpha_5(M_{32})$ which,
according to Eq.~(\ref{eq:U(1)}), also implies $\alpha_1'\le\alpha_5(M_{32})$.
The maximum
possible value of $M_{32}$, namely $M^{\rm max}_{32}$, is obtained
when $\alpha_1'=\alpha_5(M_{32})$ and is given by
the following relation obtained from Eq.~(\ref{eq:RGEY}):\\
\begin{equation}
{1\over\alpha_Y}-{1\over\alpha_5}={b_Y\over2\pi}\,
\ln{M^{\rm max}_{32}\over M_Z}\ ,
\label{eq:Mumax}
\end{equation}
which coincides with the unification scale in minimal SU(5).
We thus obtain a
set of three equations (\ref{eq:RGE2},\ref{eq:RGE3},\ref{eq:Mumax})
involving six variables:
$\alpha_2,\alpha_3, \alpha_Y$ and $\alpha_5,M^{\rm max}_{32},M_{32}$.
Solving these equations for the value of $\alpha_s(M_Z)$,
the analogue of Eq.~(\ref{eq:LO}) is given by
\begin{equation}
\alpha_s(M_Z)={\coeff{7}{3}\,\alpha\over 5\sin^2\theta_W-1
+{11\over2\pi}\,\alpha\ln(M^{\rm max}_{32}/M_{32})}\ .
\label{eq:fLO}
\end{equation}
It is clear that the prediction for $\alpha_s(M_Z)$ in flipped
SU(5) is automatically smaller than in minimal SU(5).\\

The next-to-leading order corrections to
Eq.~(\ref{eq:fLO}) are also obtained
by the substitution in Eq.~(\ref{eq:NLO}).
Numerically, an increase
of $\sim10\%$ in the denominator in Eq.~(\ref{eq:LO}),
which would compensate for
the decrease due to $\delta_{\rm 2loop}$, is achieved simply by setting
$M_{32}\approx{1\over3}M^{\rm max}_{32}$ in Eq.~(\ref{eq:fLO}).
We also note that the value of
$\alpha_5$ at $M_{32}$ is given by
\begin{equation}
{1\over\alpha_5}={1\over\alpha^{\rm max}_5}+\coeff{33}{28}\coeff{1}{2\pi}
\ln{M^{\rm max}_{32}\over M_{32}}\ ,
\label{eq:a5}
\end{equation}
where $\alpha^{\rm max}_5$ is the maximum possible value of $\alpha_5$,
which is that attained in the minimal SU(5) case.\\

The above qualitative discussion shows the differences between
flipped SU(5) and minimal SU(5). We now perform a detailed numerical
calculation of $\delta_{\rm 2loop}$, $\delta_{\rm light}$
and $M^{\rm max}_{32}$,
using the approach of
Refs.~\cite{EKNIII,ACPZ}.
The $\delta_{\rm light}$ correction is given in the
step-function approximation by
\begin{eqnarray}
\delta_{\rm light}&=&{\alpha\over20\pi}\Bigl[
-3L(m_t)+\coeff{28}{3}L(m_{\tilde g})-\coeff{32}{3}L(m_{\tilde w})
-L(m_h)-4L(m_H)\nonumber\\
&&+\coeff{5}{2}L(m_{\tilde q})-3L(m_{\tilde \ell_L})
+2L(m_{\tilde \ell_R})-\coeff{35}{36}L(m_{\tilde t_1})
-\coeff{19}{36}L(m_{\tilde t_2})\Bigr]\ ,
\label{eq:delta_l}
\end{eqnarray}
where $L(x)=\ln(x/M_Z)$. In the limit where the $L(x)$ are large,
the sparticle spectrum may be parametrized approximately
in terms of the universal soft supersymmetry-breaking scalar ($m_0$)
and gaugino ($m_{1/2}$) masses: $m_{\tilde g}=2.7 m_{1/2}$,
$m_{\tilde w}=0.79 m_{1/2}$, $m^2_i=m^2_0+c^2_i m^2_{1/2}$,
with $c_{\tilde q}=6$, $c_{\tilde \ell_L}=0.5$,
$c_{\tilde \ell_R}=0.15$. In the large-$L$ limit
we may also consider the Higgs
spectrum as containing a lighter doublet with mass $M_Z$,
and a heavier Higgs doublet whose mass is identified with the
Higgs mixing parameter $\mu$ ($m_H=|\mu|$).
Finally, the top-squark masses
are parametrized as $m^2_{\tilde t_{1,2}}=m^2_{\tilde q}\pm \bar m^2$,
where $\bar m\le m_{\tilde q}$ accounts for the splitting
of the top-squark mass eigenstates,
and is in principle related to the trilinear scalar coupling $A$.
We scan the four-dimensional parameter space
($m_0,m_{1/2},\mu,\bar
m$) in the region below 1 TeV, but
do not attempt to incorporate either the radiative
electroweak breaking constraint or any fine-tuning condition,
which would in any case lead to correlations
among the mass parameters,
and thus to a more restrictive range of possible values of
$\delta_{\rm light}$.\\

The effect ($\delta_{\rm heavy}$) of the GUT thresholds is well
known in minimal SU(5), where it yields the expression \cite{EKNIII}
\begin{equation}
\delta_{\rm heavy}^{\rm SU(5)}={\alpha\over20\pi}
\left[ -6\ln{M_U\over M_{H_3}}+4\ln{M_U\over M_V}+2\ln{M_U\over
M_\Sigma}\right]\ ,
\label{heavySU5}
\end{equation}
where $M_U={\rm max}\{M_{H_3},M_V,M_\Sigma\}$. In this expression, $M_V$
represents the mass of the $X,Y$ gauge bosons and gauginos, and the
Higgs bosons and Higgsinos in the \r{24};
$M_\Sigma$ represents the mass of the
remaining states in the \r{24} (a color octet and an SU(2) triplet);
and $M_{H_3}$ represents the common
mass of the triplet components of the Higgs pentaplets.
Since $M_{H_3}>M_V$ is required to suppress
dimension-five proton decay operators in minimal SU(5) \cite{AN,LNP},
$\delta_{\rm heavy}^{\rm SU(5)}>0$ in general \cite{EKNIII},
unless one resorts to unnaturally large sparticle masses.\\

The corresponding expression for $\delta_{\rm heavy}$
in flipped SU(5) can be
obtained from the general formula \cite{EKNIII}
\begin{equation}
\delta_{\rm heavy}=-{\alpha\over20\pi}\sum_{R_i}
C(R_i)\,\ln{M_U\over M_i}\ ,
\label{deltahg}
\end{equation}
where the sum runs over all GUT scale particles $R_i$
with masses $M_i$, and
\begin{equation}
C(R)=\coeff{10}{3}b_Y(R)-8b_2(R)+\coeff{14}{3}b_3(R)\ ,
\label{C}
\end{equation}
is a linear combination of the one-loop contributions
to the beta functions.
The spectrum of heavy particles in flipped SU(5)
consists of the $X,Y$ gauge supermultiplets,
the \r{10} and \rb{10} Higgs supermultiplets
$H=\{Q_H,d^c_H,\nu^c_H\}$ and
$\bar H=\{Q_{\bar H},d^c_{\bar H},\nu^c_{\bar H}\}$
that break SU(5), and the Higgs triplets ($h_3,\bar h_3$)
contained in the Higgs pentaplets ($h,\bar h$).
\footnote{There are also several gauge-singlet
representations, such as the right-handed neutrinos,
which have large
masses but do not contribute to $\delta_{\rm heavy}$,
since they have $C(R) = 0$.}
In the SU(5) symmetry breaking process,
the neutral components of the
\r{10} and \rb{10} Higgs multiplets
acquire equal vevs $\vev{\nu^c_H}=\vev{\nu^c_{\bar H}}=V$, giving
equal masses ($M_V=g_5|V|$) to the $X,Y$ gauge bosons and gauginos,
and to the $Q_H,Q_{\bar H}$ Higgs boson and Higgsinos.
The remaining components ($d^c_H,d^c_{\bar H}$)
of the \r{10} and \rb{10} participate in the economical
missing partner mechanism of flipped SU(5), combining with the
$h_3,\bar h_3$ triplets in the Higgs pentaplets ($h,\bar h$) via the
superpotential couplings $\lambda_4\,HHh$
and $\lambda_5\,\bar H\bar H\bar h$ to
acquire masses $M_{H_3}=\lambda_4|V|$
and $M_{\bar H_3}=\lambda_5|V|$. We note that,
unlike the case of minimal SU(5), there are no contributions from
$\Sigma$-type fields since these are absent from the spectrum of the
model. (We recall that the $\nu^c_H$ and $\nu^c_{\bar H}$ are gauge
singlets, and hence do not contribute to $\delta_{\rm heavy}$).
Calculating the
$C(R)$ coefficients in Eq.~(\ref{deltahg}), we obtain
\begin{equation}
\delta_{\rm heavy}={\alpha\over20\pi}
\left[ -6\ln{M_{32}\over M_{H_3}}-6\ln{M_{32}\over M_{\bar H_3}}
+4\ln{M_{32}\over M_V}\right]={\alpha\over20\pi}
\left[ -6\ln{r^{4/3}g^{2/3}_5\over\lambda_4\lambda_5}\right]
\label{heavyfSU5}
\end{equation}
where $r={\rm max}\{g_5,\lambda_4,\lambda_5\}$.
Since the $H_3$ and $\bar H_3$ do not mix, they do not contribute
significantly to proton decay, and hence there is no strong
constraint on $M_{H_3,\bar H_3}$ from proton decay in
flipped SU(5). We therefore see from Eq.~(\ref{heavyfSU5})
that it is possible that $M_{H_3,\bar H_3}<M_V=M_U$
(\ie, $r=g_5$), in which case
$\delta_{\rm heavy}<0$ naturally. For instance, if
$\lambda_4,\lambda_5\sim{1\over8}g_5$,
then $\delta_{\rm heavy}\approx-0.0030$,
which completely compensates the $\delta_{\rm 2loop}$ contribution.\\

\begin{figure}[p]
\vspace{6.5in}
\includegraphics{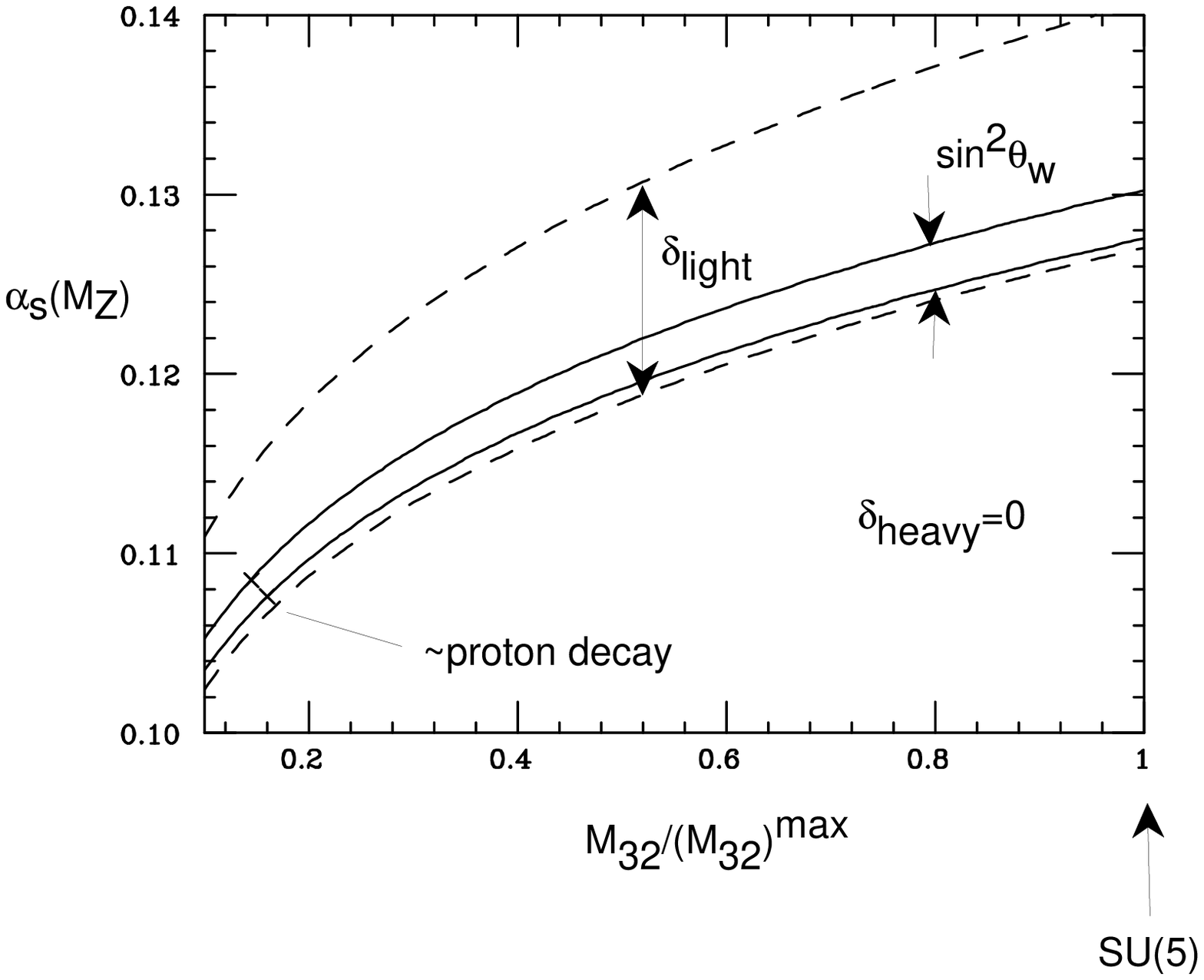}
\caption{The prediction for $\alpha_s(M_Z)$ in flipped SU(5) as a function
of the ratio $M_{32}/M^{\rm max}_{32}$. The minimal SU(5) predictions are
recovered for $M_{32}/M^{\rm max}_{32}=1$. The solid lines represent the range
of predictions
for $\sin^2\theta_W=0.23143\pm0.00028$ with no threshold corrections
($\delta_{\rm light}=\delta_{\rm heavy}=0$). The dashed lines represent the
excursion obtained by scanning over sparticle masses ($\delta_{\rm
light}\not=0$) below 1 TeV. Inclusion of typical $\delta_{\rm heavy}$ values
amounts to a rescaling of the $M_{32}/M^{\rm max}_{32}$ axis by a factor in the
range 0.8--1.8. The crosses indicate approximate lower bounds from proton decay
constraints, which are displayed in more detail in Figure~2.}
\label{fig:sin2r}
\end{figure}

In Fig.~\ref{fig:sin2r} we show the prediction for
$\alpha_s(M_Z)$ in flipped
SU(5) as a function of the ratio $M_{32}/M^{\rm max}_{32}$.
The solid lines delimit
the range of predictions obtained using the latest experimental value
$\sin^2\theta_W=0.23143\pm0.00028$ and
setting $\delta_{\rm light}+\delta_{\rm heavy}=0$.
We note that the minimal SU(5) prediction
is obtained when $M_{32}/M^{\rm max}_{32}=1$.
Scanning over the possible range of sparticle masses
described above, we find significant variations in the
predictions for $\alpha_s(M_Z)$, indicated by the
dashed lines in Fig.~\ref{fig:sin2r}, usually
towards higher values (\ie, $\delta_{\rm light}>0$).
The maximum values of
$\alpha_s(M_Z)$ are obtained for $m_0\gg m_{1/2},|\mu|$,
whereas the minimum
values are obtained for $|\mu|\gg m_0,m_{1/2}$.
\footnote{Note, however, that this scenario is
disfavored by radiative symmetry-breaking constraints,
which imply $|\mu|\sim{\rm max}\,\{m_0,m_{1/2}\}$.}
The general detrimental effect of $\delta_{\rm light}$
may be compensated somewhat by $\delta_{\rm heavy}<0$ effects.\\

In the case of minimal SU(5) (\ie, $M_{32}/M^{\rm max}_{32}=1$),
these effects may decrease the
value of $\alpha_s(M_Z)$ as low as $0.123$ \cite{baggeretal}.
However, in flipped SU(5) one may find
$\alpha_s(M_Z)\lsim0.120$ even without resorting to such strategems,
if the unification scale is lowered enough.
Equations (\ref{eq:NLO}) and (\ref{eq:fLO}) show that
including the effects of
$\delta_{\rm heavy}$ simply amounts to a re-scaling
of the $M_{32}/M^{\rm max}_{32}$
axis on Fig.~\ref{fig:sin2r}, \ie,
\begin{equation}
{M_{32}\over M^{\rm max}_{32}}\to {M_{32}\over M^{\rm max}_{32}}
\ e^{-10\pi\,\delta_{\rm heavy}/11\alpha}\ .
\label{scaling}
\end{equation}
For definiteness, let us assume that ${1\over \rho}\lsim
g_5/\sqrt{\lambda_4\lambda_5}\lsim \rho$,
with the heuristic choice $\rho=3$.
Equation~(\ref{heavyfSU5}) then implies $0.0005\gsim\delta_{\rm
heavy}\gsim-0.0016$, and the multiplicative factor in Eq.~(\ref{scaling})
is in the range $0.81\lsim e^{-10\pi\,
\delta_{\rm heavy}/11\alpha}\lsim1.8$.\\

We now turn to a possible experimental signature of this
flipped SU(5) mechanism for lowering $\alpha_s$, namely
proton decay, which is clearly enhanced by
decreasing the unification scale.
As is well known, one of the attractive features of flipped SU(5)
is the fact that dimension-five proton decay operators are
highly suppressed in this model \cite{faspects}. However, the usual
dimension-six proton decay operators which would lead, for
example, to $p\to e^+\pi^0$ decay, are present and
their coefficients would be enhanced
relative to those in minimal SU(5) by a factor of
$(M^{\rm max}_{32}/M_{32})^4$. Concretely, we estimate that
\cite{faspects,HMY}
\begin{equation}
\tau(p\to e^+\pi^0)\approx1.5\times10^{33}
\left({M_{32}\over10^{15}\GeV}\right)^4
\left({0.042\over\alpha_5}\right)^2 ,
\label{eq:taupf}
\end{equation}
where we have taken the central
value of the $p \rightarrow e^+\pi^0$ matrix element from
\cite{GKSMPT}, which quotes an error of about $20 \%$ in the
decay rate.
The error quoted in \cite{GKSMPT} is mainly statistical in
nature, and does not include possible systematic errors
associated with the quenched-fermion approximation, finite lattice
size and spacing, etc. We consider it prudent to double the
quoted error when interpreting (\ref{eq:taupf}) in order to
allow for such possible effects. This would,
however, correspond to a variation in $M_{32}$
of no more than $10 \%$.\\

\begin{figure}[p]
\vspace{6.5in}
\includegraphics{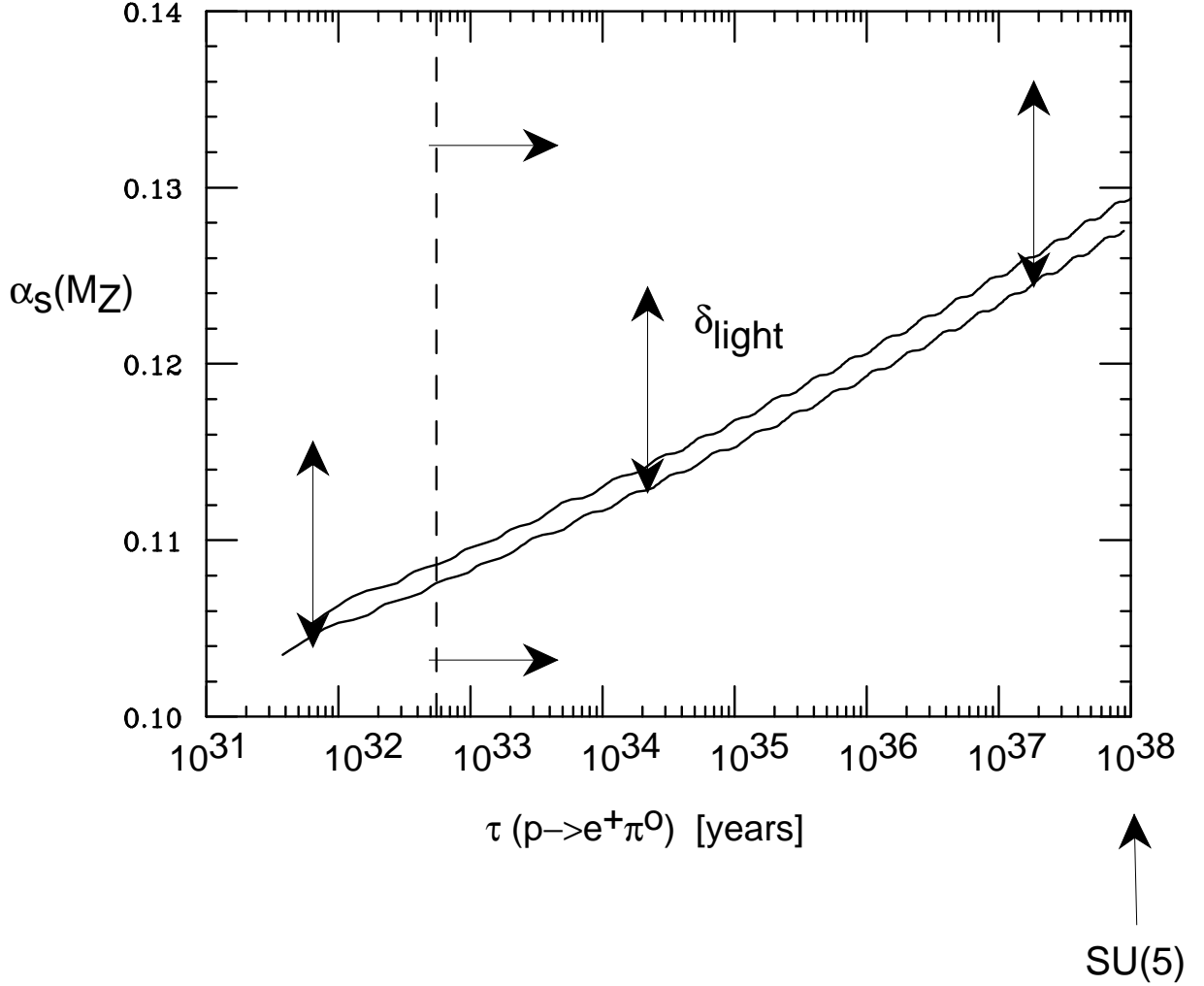}
\caption{The prediction for $\alpha_s(M_Z)$ in flipped SU(5) as a function
of the proton lifetime into the mode $e^+\pi^0$. The present experimental lower
bound is indicated by the dashed line. The minimal SU(5) predictions are
recovered for $M_{32}/M^{\rm max}_{32}=1$. The solid lines represented the
range of predictions for $\sin^2\theta_W=0.23143\pm0.00028$ with no threshold
corrections ($\delta_{\rm light}=\delta_{\rm heavy}=0$).
The arrows indicate
the extent of the excursion obtained by scanning over sparticle masses
($\delta_{\rm light}\not=0$) below 1 TeV. Heavy threshold corrections amount
to a multiplicative uncertainty factor in the proton lifetime in the range
0.4--10.}
\label{fig:taup}
\end{figure}

Equipped with equation (\ref{eq:taupf}), we plot the prediction for
$\alpha_s(M_Z)$ versus $\tau(p\to e^+\pi^0)$ in
Fig.~\ref{fig:taup}, indicating
the effect of $\delta_{\rm light}$.
As discussed above, for a fixed value of $\alpha_s(M_Z)$,
heavy threshold corrections introduce a multiplicative
uncertainty in $M_{32}$ in the range
0.8--1.8, and thus a multiplicative uncertainty
in the proton lifetime in the range 0.4--10. It is
interesting to note that the present experimental
lower bound $\tau(p\to e^+\pi^0)^{\rm exp}>5.5\times10^{32}\y$ \cite{PDG}
allows values of $\alpha_s(M_Z)$ as low as 0.108,
even without accounting for
the uncertainties in the proton decay calculation.
This approximate lower
bound is indicated by the crosses in Fig.~\ref{fig:sin2r}. Moreover,
for values of $\alpha_s(M_Z)\lsim0.114$,
the mode $p\to e^+\pi^0$
should be observable at SuperKamiokande in flipped SU(5), whereas
it is expected to be utterly
unobservable in minimal supersymmetric
SU(5). Furthermore, even though flipped
SU(5) and minimal supersymmetric
SU(5) predict similar proton decay modes,
the relative branching ratios differ, because of the different
operator structures in the two models. We also recall that
the mode $p\to\bar\nu K^+$ via dimension-five operators
should be dominant in minimal supersymmetric SU(5) \cite{AN}.\\

We conclude by putting the present investigation in the
proper perspective regarding flipped SU(5) model building.
As is well known, thanks to the
simple ``flip" $e^c\leftrightarrow\nu^c$, flipped SU(5) can be broken
by vevs of small representations (\r{10},\rb{10}) in a unique way. The
presence of $\nu^c$ leads to a see-saw mechanism for generating small
neutrino masses, and possibly
a cosmological baryon asymmetry which is later recycled
into a baryon asymmetry by sphaleron interactions \cite{FY}.
The triplet components of
the \r{10},\rb{10} yield a very economical doublet-triplet
splitting, which in turn suppresses greatly dimension-five proton decay
operators. This natural suppression allows the heavy Higgs triplets to
contribute negatively to $\delta_{\rm heavy}$,
so that the prediction for $\alpha_s(M_Z)$ can
be lowered to anywhere in the 0.11-0.12 interval. If in the future
the low- and high-energy determinations
of $\alpha_s(M_Z)$ are reconciled
experimentally with a value in the lower half of the range
presently allowed, the SuperKamiokande detector
should observe $p\to e^+\pi^0$ decays, whereas minimal
supersymmetric SU(5) would predict $p\to\bar\nu K^+$.

\section*{Acknowledgments}
The work of J.~L. has been supported in part by
DOE grant DE-FG05-93-ER-40717.
The work of D.V.N. has been supported in part by
DOE grant DE-FG05-91-ER-40633.

\newpage

\clearpage

\clearpage

\end{document}